\documentclass[aps,prc,reprint,amsmath,amssymb,showpacs,superscriptaddress]{revtex4-1}

\usepackage{CJK}
\usepackage{graphicx}
\usepackage{dcolumn}
\usepackage{bm}
\usepackage{epstopdf}

\usepackage{color}
\usepackage{hyperref}

\usepackage{threeparttable}
\usepackage{booktabs}
\usepackage{multirow}
\usepackage{longtable}
\usepackage{supertabular}

\begin{document}
\begin{CJK*}{UTF8}{} 


\title{Nuclear Mass Predictions Using a Neural Network with Additive Gaussian Process Regression-Optimized Activation Functions}

\author{H. X. Liu \CJKfamily{gbsn} (刘辉鑫)}
\affiliation{Department of Physics, Fuzhou University, Fuzhou 350108, Fujian, China}


\author{S. Manzhos}
\affiliation{School of Materials and Chemical Technology, Institute of Science Tokyo, Tokyo 152-8552, Japan}

\author{X. H. Wu \CJKfamily{gbsn} (吴鑫辉)}
\email{wuxinhui@fzu.edu.cn}
\affiliation{Department of Physics, Fuzhou University, Fuzhou 350108, Fujian, China}


\begin{abstract}
  Nuclear masses are machine-learned as a function of proton and neutron numbers.
  The neural network with additive Gaussian process regression-optimized activation functions (GPR-NN) method is employed for the first time for this purpose.
  GPR-NN combines the advantages of both neural networks and Gaussian process regression, in that it possesses the expressive power of an NN, in principle allowing modeling any kind of dependence of nuclear mass on the features, and robustness of a linear regression with respect to overfitting.
  A study of the GPR-NN approach for interpolation and extrapolation in nuclear mass predictions is presented.
  It is found that the optimal hyperparameters for the GPR-NN approach in interpolation and extrapolation are different.
  If an appropriate set of hyperparameters is adopted, the GPR-NN approach can achieve good extrapolation performance for nuclear mass prediction, which could potentially help improve the mass predictions of a large number of currently experimentally unknown nuclei.  
\end{abstract}

\maketitle

\end{CJK*}


\section{Introduction}

The knowledge of nuclear masses is important for not only nuclear physics~\cite{Yamaguchi2021PPNP, Monteagudo2024PRL, Zhang2025AAPPSBulletin}, but also nuclear astrophysics~\cite{Mumpower2016Prog.Part.Nucl.Phys., Jiang2021Astrophys.J., Wu2023Sci.Bull.}.
Experimentally, about 2500 nuclear masses have been measured to date~\cite{Wang2021Chin.Phys.C}.
Nevertheless, most exotic nuclei still remain beyond the current experimental capabilities, especially the neutron-rich ones related to the $r$-process nucleosynthesis.
Therefore, nuclear mass predictions from models are essential.
Many efforts have been made to describe nuclear masses, including macroscopic models~\cite{Weizsaecker1935Z.Physik}, 
macroscopic-microscopic models~\cite{Pearson1996Phys.Lett.B, Wang2014Phys.Lett.B, Moeller2016Atom.DataNucl.DataTables, Koura2005Prog.Theor.Phys.},
and microscopic models~\cite{Goriely2009Phys.Rev.Lett., Goriely2009Phys.Rev.Lett.a, Pena2016EPJA, Xia2018Atom.DataNucl.DataTables, Yang2021Phys.Rev.C, Zhang2022Atom.DataNucl.DataTables}.
Among these models, the macroscopic-microscopic models, e.g., FRDM~\cite{Moeller2016Atom.DataNucl.DataTables} and WS4~\cite{Wang2014Phys.Lett.B}, are the ones most frequently used in related studies, especially in the $r$-process studies~\cite{Cyburt2010Astrophys.J.Suppl.Ser., Zhao2019Astrophys.J., Jiang2021Astrophys.J., Wu2022Astrophys.J.152, Huang2025ApJ}.
However, different models can predict very different nuclear masses for neutron-rich exotic nuclei far away from the experimentally known region, indicating that uncertainties of theoretical models are still quite significant in predicting these exotic nuclei.

Efforts in two directions are ongoing to provide accurate nuclear mass predictions. 
One direction is to build microscopic nuclear mass models with more effects being included, e.g., the on-going DRHBc mass table project~\cite{Zhang2022Atom.DataNucl.DataTables, Pan2022Phys.Rev.C, Guo2024ADNDT} which includes pairing, deformation, and continuum effects simultaneously.
Microscopic models are usually believed to have a better reliability of extrapolation~\cite{Zhao2012Phys.Rev.C, Zhang2025AAPPSBulletin}, although their precision of pre-dicting experimentally known masses is currently poorer than that of the macroscopic-microscopic models.

Meanwhile, the other direction involves improving nuclear mass predictions using machine learning techniques~\cite{Utama2016Phys.Rev.C, Neufcourt2018Phys.Rev.C, Niu2018Phys.Lett.B, Neufcourt2019Phys.Rev.Lett., Niu2022Phys.Rev.C, Li2024PLB, Wu2020Phys.Rev.C051301, Du2023Chin.Phys.C, Wu2024Phys.Rev.C, Guo2024PRC}.
The problem of nuclear mass prediction can be said to have become a testbed for the application of machine learning in nuclear physics.
Most standard machine-learning approaches have been employed in nuclear mass studies, such as the kernel ridge regression (KRR)~\cite{ Wu2020Phys.Rev.C051301, Wu2021Phys.Lett.B, Wu2022Phys.Lett.B137394, Wu2024PRC_AKRR} and Gaussian process regression (GPR) approaches~\cite{Neufcourt2019Phys.Rev.Lett.}, the radial basis function (RBF) approach~\cite{Wang2011Phys.Rev.C, Niu2016Phys.Rev.C}, the (Bayesian) neural network (NN) approach~\cite{Utama2016Phys.Rev.C, Niu2018Phys.Lett.B},  the principal component analysis (PCA) approach~\cite{Wu2024SC, Giuliani2024PRR}, and so on.
After a machine learning approach is successfully applied to nuclear mass, it will be promoted to applications in other aspects of nuclear physics.
For example, the successful applications of the KRR approach in nuclear masses~\cite{Wu2020Phys.Rev.C051301, Wu2021Phys.Lett.B, Wu2022Phys.Lett.B137394,  Du2023Chin.Phys.C, Wu2024Phys.Rev.C, Wu2024PRC_AKRR, Guo2024PRC, Guo2022Symmetry, Wu2023Front.Phys.} have also stimulated its applications to other topics in nuclear physics, including the energy density functionals~\cite{Wu2022Phys.Rev.C, Chen2024IJMPE, Wu2025CP}, charge radii~\cite{Ma2022Chin.Phys.C, Tang2024NST}, and neutron-capture cross-sections~\cite{Huang2022Commun.Theor.Phys.}.

Different machine-learning approaches have different advantages and disadvantages.
For example, the neural network (NN) approach~\cite{Montavon2012neural} has the advantage of high expressive power (universal approximator property) but requires non-linear optimization of a large number of parameters, which exacerbates the problem of overfitting and local minima, and can be CPU-intense for large NNs. 
Kernel regressions~\cite{Bishop2006pattern} such as the Gaussian process regression (GPR) combine the expressive power of a nonlinear method (achieved with nonlinear kernels) and robustness of regularized linear regression that it is; this can provide reliable machine learning from small datasets, and interpretability, especially with the use of structured kernels~\cite{Thant2025CPR, Manzhos2023AC}, but kernel regression is difficult to use with large datasets and high-dimensional kernels~\cite{Candela2005JMLR, Manzhos2024JCP, manzhos2023JCP}. The expressive power is also limited by the choices of nonlinear kernels that can only capture a restricted range of complex relationships. 

A recently proposed machine learning approach, i.e., neural network with additive Gaussian process regression-optimized activation functions (GPR-NN)~\cite{Manzhos2023JPCA}, combines the advantages of both NN and GPR approaches. It builds a representation of the target function that has the same form as a single hidden layer NN with optimal shapes of neuron activation functions in the feature space, while algorithmically it is 1st order additive GPR in the space of redundant coordinates corresponding to neuron arguments. The use of additive kernels avoids problems associated with multidimensional kernels. As the redundant coordinates (corresponding to NN weights matrix) are defined by rules and no nonlinear optimization is performed, the method avoids overfitting as the number of neurons is grown beyond optimal, and obviates the problem of local minima. All neuron activation functions are optimal for given data and given redundant coordinates (weight matrix) and are obtained in one linear step.

The GPR-NN approach has been successfully employed to the construction of molecular potential energy surfaces~\cite{Manzhos2023JPCA, Manzhos2023JCP1}, to predicting properties of materials from chemical composition and structure (materials informatics)~\cite{Tang2025JMI, Nukunudompanich2024MRS}, and to neuromorphic computing~\cite{Manzhos2024JPCT}. 
While being general, it allows interpretative ML including analysis of feature importance and of the type of functional dependence of the target on the features~\cite{Tang2025JMI, Manzhos2025MLST}.

In this work, the GPR-NN approach is employed to improve the nuclear mass predictions.
The corresponding hyperparameters, including number of redundant coordinates $R$, the length scale of the kernel $L$, and the regularization parameter $\delta$, are studied and optimized through careful validations for both interpolation and extrapolation.
The performance and reliability of the GPR-NN approach in extrapolating nuclear mass predictions are analyzed in detail.

\section{Theoretical framework}

The GPR-NN approach~\cite{Manzhos2023JPCA} is a hybrid between a single-hidden layer NN and GPR. 
The target function $f({\bm x})$, ${\bm x}\in R^d$ is represented as a first-order additive model in redundant coordinates ${\bm y}\in R^D$, $D>d$, with the component functions constructed with GPR:
\begin{equation}\label{Eq:gprnn}
\begin{aligned}
  f({\bm x}) = & \sum_{n=1}^D f_n(y_n) = \sum_{n=1}^{D}\left[\sum_{m=1}^{M}k(y_n,y_n^{(m)})c_m\right] \\
  & = \sum_{m=1}^{M}\left[\sum_{n=1}^{D}k(y_n,y_n^{(m)})\right]c_m,
\end{aligned}
\end{equation}
where $k(y_n,y_n^{(m)})$ is the kernel, and $m$ indexes training data points. 
Each component function $f_n(y_n)$ and the corresponding kernel are one-dimensional and therefore issues with (non-additive) high-dimensional kernels are avoided.
The coefficients ${\bm c}$ are obtained  using standard GPR methodology:
\begin{equation}
  {\bm c} = ({\bm K}+\delta {\bm I})^{-1}{\bm f},
\end{equation}
with
\begin{equation}
  \bm{K} =
\begin{pmatrix}
  k(\bm{y}^{(1)}, \bm{y}^{(1)})  & k(\bm{y}^{(1)}, \bm{y}^{(2)}) & \cdots & k(\bm{y}^{(1)}, \bm{y}^{(M)}) \\
  k(\bm{y}^{(2)}, \bm{y}^{(1)}) & k(\bm{y}^{(2)}, \bm{y}^{(2)})  & \cdots & k(\bm{y}^{(2)}, \bm{y}^{(M)}) \\
  \vdots & \vdots & \ddots & \vdots \\
  k(\bm{y}^{(M)}, \bm{y}^{(1)}) & k(\bm{y}^{(M)}, \bm{y}^{(2)}) & \cdots & k(\bm{y}^{(M)}, \bm{y}^{(M)}) 
\end{pmatrix}
\end{equation}
where $\delta$ is the regularization parameter and $k(\bm{y}^{(m)},\bm{y}^{(m')})$ is the additive kernel function that is specified as
\begin{equation}
\begin{aligned}
  k(\bm{y}^{(m)},\bm{y}^{(m')})= & \sum_{n=1}^{D}k(y_n^{(m)},y_n^{(m'}) \\
  & =\sum_{n=1}^{D}\exp\left[-\frac{1}{2L^2}(y_n^{(m)}-y^{(m')}_n)^2\right],
\end{aligned}
\end{equation}
where $L$ represents the length parameter of the kernel.

The redundant coordinates $y_n$ are linear functions of ${\bm x}$, ${\bm y} = {\bm W}{\bm x}$, where ${\bm W}$ is defined by rules and is not optimized. 
Here, the original coordinates ${\bm x}$ are included as a subset of ${\bm y}$ ($y_n=x_n$ for $n=1,2,...,d$) and the rows of matrix ${\bm W}$ defining other $y_n$ ($n>d$) are chosen as elements of a $d$-dimensional Sobol sequence~\cite{Sobol1967USSR}.
Control of the number of redundant coordinates ($R=D-d$) allows studying important hidden features and the coupling of features.
All terms $f_n(y_n)$ of Eq.~\eqref{Eq:gprnn} are computed as a single linear step with a standard GPR code. 
The only computational cost overhead vs. standard GPR is the summation in the kernel. 
Functions $f_n(y_n)$ are optimal in the least squares sense for given ${\bm W}$ and data.

\begin{figure}[!htbt]
  \centering
  \includegraphics[width=0.45\textwidth]{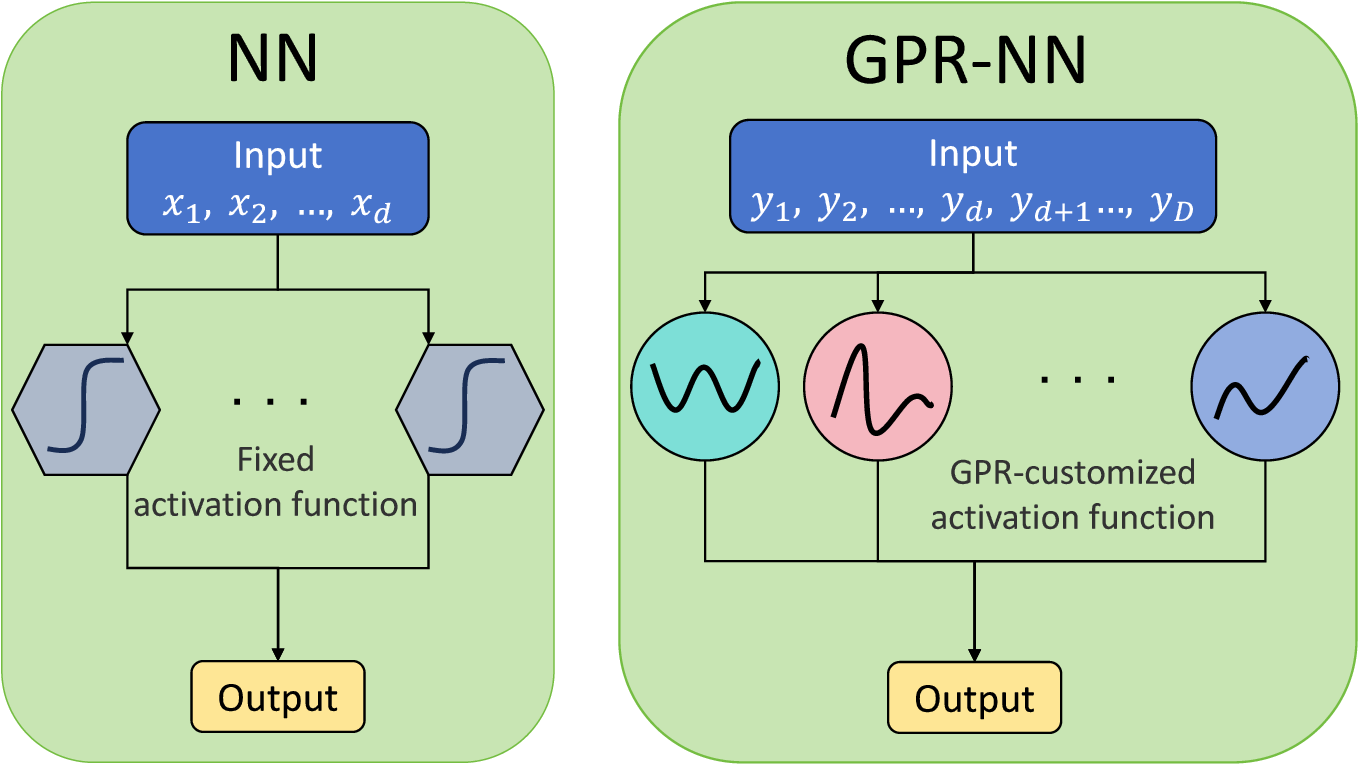}
  \caption{Schematic diagram of the GPR-NN approach--A comparison of NN and GPR-NN.}
\label{fig1}
\end{figure}

The schematic diagram of the GPR-NN approach is illustrated in Fig.~\ref{fig1}.
In the space of the original features ${\bm x}$, it is analogous to to a single-hidden layer NN with $D$ neurons, with optimal shapes of neuron activation functions for each neuron.
It possesses therefore a universal approximator property. 
Note that biases and output weights are subsumed in the definition of $f_n(y_n)$ and need not be considered separately. 
As no nonlinear optimization is done, the method is as robust as linear regression (as GPR is a regularized linear regression with nonlinear basis functions derived from the kernel function), and there is no overfitting as $D$ exceed the optimal number of neurons.

\section{Numerical details}

The calculations are performed based on the MATLAB code of the GPR-NN approach developed in Ref.~\cite{Manzhos2023JPCA}.
Applying the GPR-NN approach to the studies of nuclear mass predictions, the input is proton number and neutron number, i.e., ${\bm x}=(Z,N)$.
After the redundant coordinates $y_n$ are generated with the Sobol sequence, the $y_n$ for all the training nuclei are scaled to [0, 1].
The target function is nuclear mass residuals $M_{\rm res}$, i.e., deviations between experimental data and theoretical predictions.
Therefore, the predicted mass for a nucleus $(Z, N)$ is, thus, given by $M_{\rm GPR-NN}(Z, N)=M_{\rm th}(Z, N)+M_{\rm res}(Z, N)$.

In the training process of the GPR-NN, the experimental nuclear masses from AME2020~\cite{Wang2021Chin.Phys.C} are taken for the nuclei with $Z\geq8$ and $N\geq8$, while the masses with experimental error exceeding 100 keV are excluded.
The theoretical predictions are taken from the RCHB nuclear mass model~\cite{Xia2018Atom.DataNucl.DataTables}.
An overlap between the processed AME2020 and RCHB mass table was retained, encompassing 2278 nuclei.
During the process of validating generalization ability, experimental mass data sets, AME1983~\cite{Wapstra1985NPA}, AME1993~\cite{Audi1993NPA}, AME2003~\cite{Audi2003NPA}, and AME2012~\cite{Wang2012Chin.phys.C} are also utilized.

\section{Results and discussion}

The hyperparameters involved in the GPR-NN approach includes the number of redundant coordinates $R$, the length scale of the kernel $L$, and the regularization parameter $\delta$.
In order to determine the hyperparameters $(R,L,\delta)$, the data set of 2278 nuclei is randomly divided into ten bins of equal size (for convenience, eight nuclei are ignored randomly in the partitioning).
For each bin, the GPR-NN is trained on the remaining samples using a series of hyperparameter sets.
The obtained rms deviations are used to evaluate the performance of the corresponding hyperparameter sets.
This procedure is also known as the so-called tenfold cross-validation, which is much efficient than the leave one out cross-validation.

\begin{figure}[!htbp]
  \centering
  \includegraphics[width=0.45\textwidth]{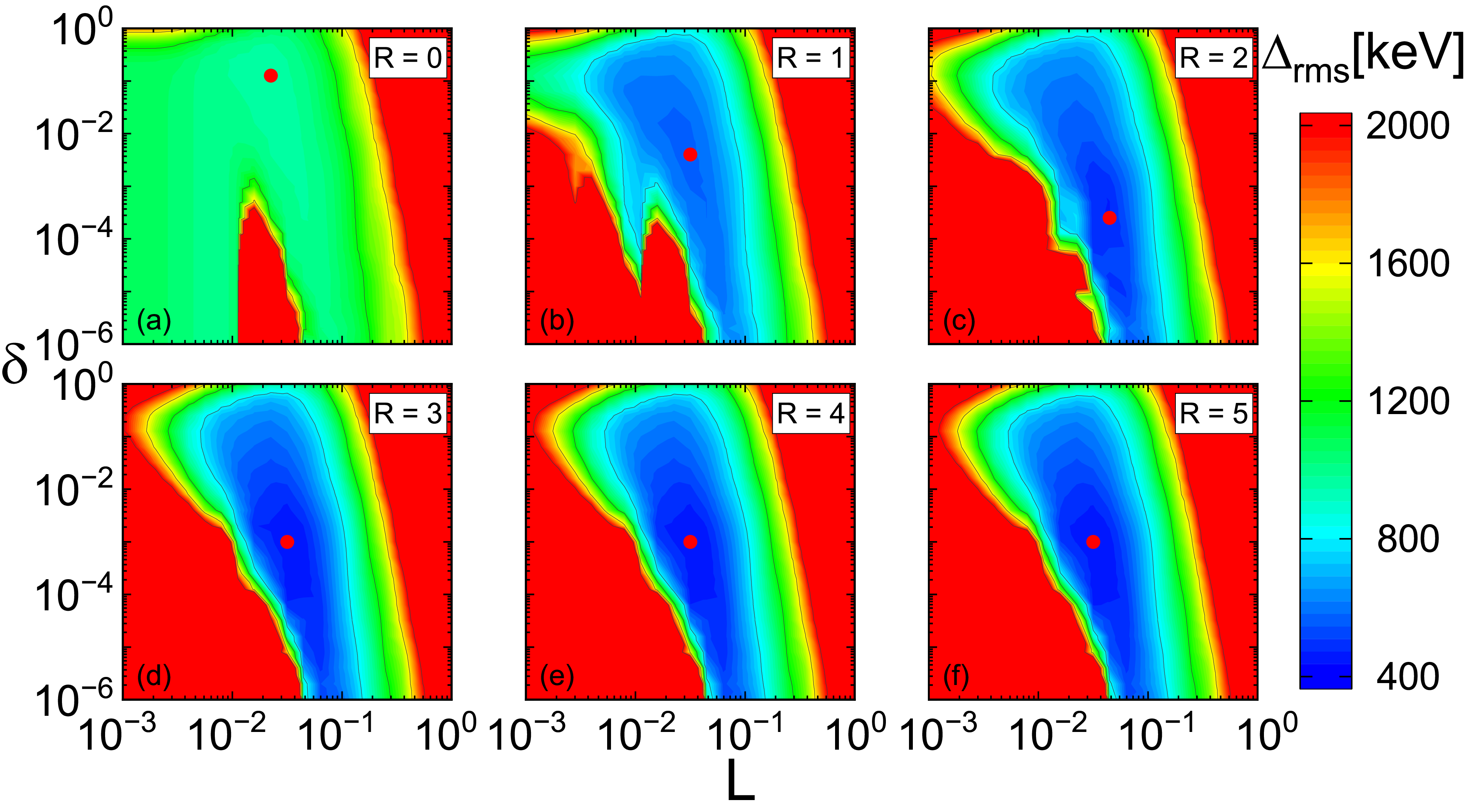}
  \caption{The rms deviations obtained by the tenfold cross-validation with different hyperparameters $(R,L,\delta)$.
  Each panel presents the results with a specific number of redundant coordinates $R$.
  The minima are labeled with red dots.}
  \label{fig2}
\end{figure}

The rms deviations $\Delta_{\rm rms}$ of the GPR-NN predictions relative to the experimental data under different sets of hyperparameters are presented in Fig.~\ref{fig2}.
As can be seen from Fig.~\ref{fig2}, the performance of the GPR-NN approach is  affected by the hyperparameters, and thus one should carefully validate the hyperparameters $(R,L,\delta)$. On the other hand, in the basin around the optimal values of $(R,L,\delta)$, the results are relatively stable with respect to specific values of $(R,L,\delta)$, showing that the method is robust. 
One can see from each subplot with fixed $R$, the hyperparameters $(L,\delta)$ can be well determined according to the minima of the rms deviations. The hyperparameters $L$ and $\delta$ do not sensitively depend on $R$, which enables increasing $R$ until the test error plateaus without or with minimal effort of retuning the hyperparameters.

\begin{figure}[!htbp]
  \centering
  \includegraphics[width=0.45\textwidth]{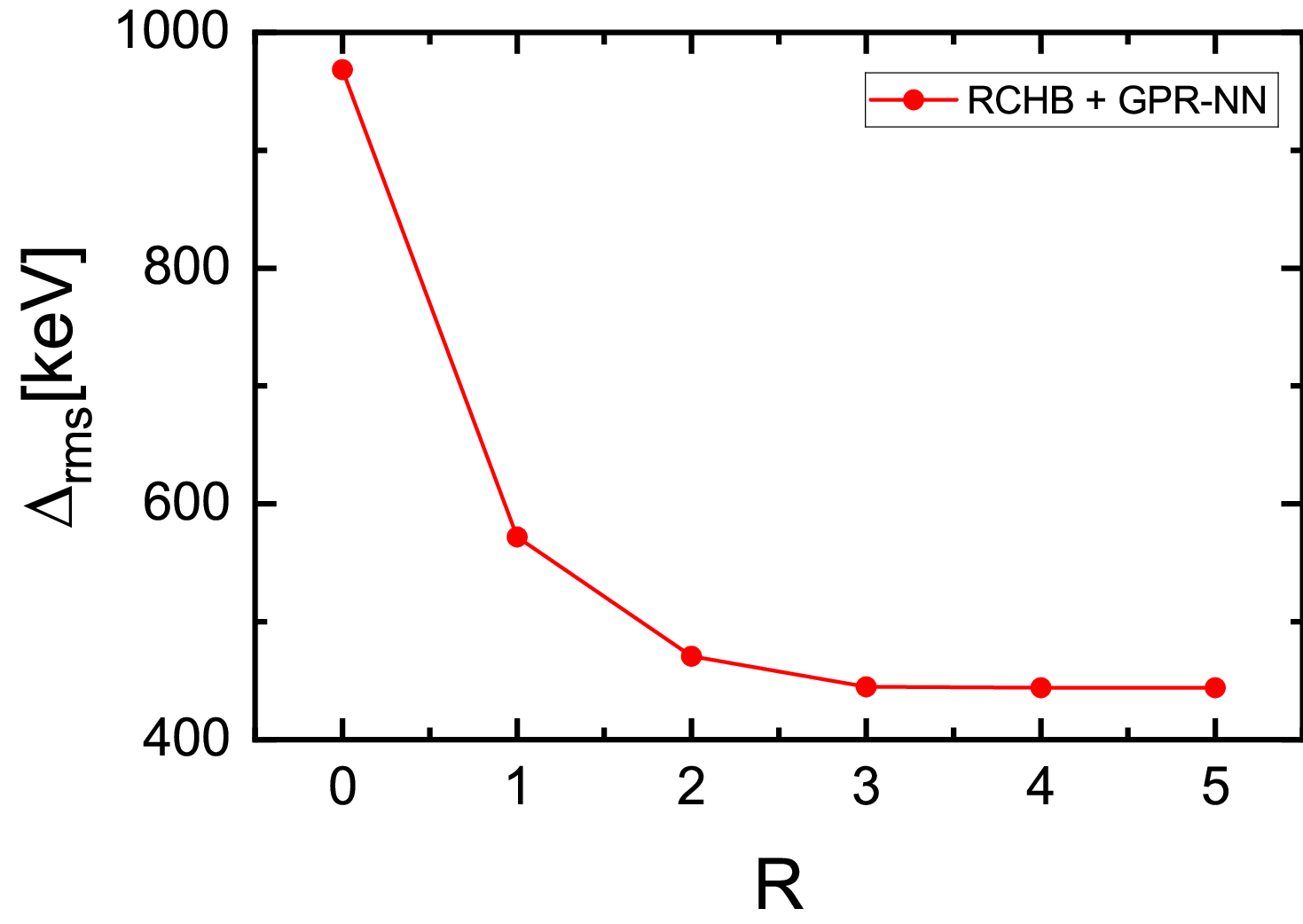}
  \caption{The minima of the rms deviations $\Delta_{\rm rms}$ obtained by the tenfold cross-validation by optimizing over $L$ and $\delta$, for each specific number of redundant coordinates $R$.
  }
  \label{fig3}
\end{figure}

It can also be noticed from Fig.~\ref{fig2} that the plots with $R\geq 3$ are generally identical, which indicates that the number of redundant coordinates $R=3$ would be enough.
This can be seen more clearly in Fig.~\ref{fig3}, where the minima of the rms deviations $\Delta_{\rm rms}$ for each given $R$ are presented.
For $R=0$, the rms deviation $\Delta_{\rm rms}$ is still as large as $\sim 1000$ keV, and then the $\Delta_{\rm rms}$ decreases with the inclusion of redundant coordinates.
This indicates the importance of coupling between $Z$ and $N$.
When $R\geq 3$, the $\Delta_{\rm rms}$ converges to about $440$~keV.
This means when applying the GPR-NN approach to the studies of nuclear masses, the number of redundant coordinates $R=3$ is typically sufficient, and this value will be adopted for the remainder of this work. Fig.~\ref{fig3} also highlights the robustness of the method in that any sufficiently high value of $R$ will achieve a similar level of error, there is no overfitting when $R$ is larger than an optimal value. 
In the case of $R=3$, the other two optimized hyperparameters obtained from the tenfold cross-validation (see Fig.~\ref{fig2}~(d)) are $(L=0.031, \delta=0.001)$.
This represents the optimized hyperparameters of the GPR-NN approach in the interpolation.

The predictive power of a machine-learning-based approach when extrapolating to experimentally unknown regions is more important.
To evaluate this, similar to Ref.~\cite{Wu2020Phys.Rev.C051301}, for each isotopic chain, the eight most neutron-rich nuclei are removed from the training set, and they are classified into eight test sets respectively, corresponding to the different extrapolation distances from the remain training set in the neutron direction.

\begin{figure}[!htbp]
  \centering
  \includegraphics[width=0.45\textwidth]{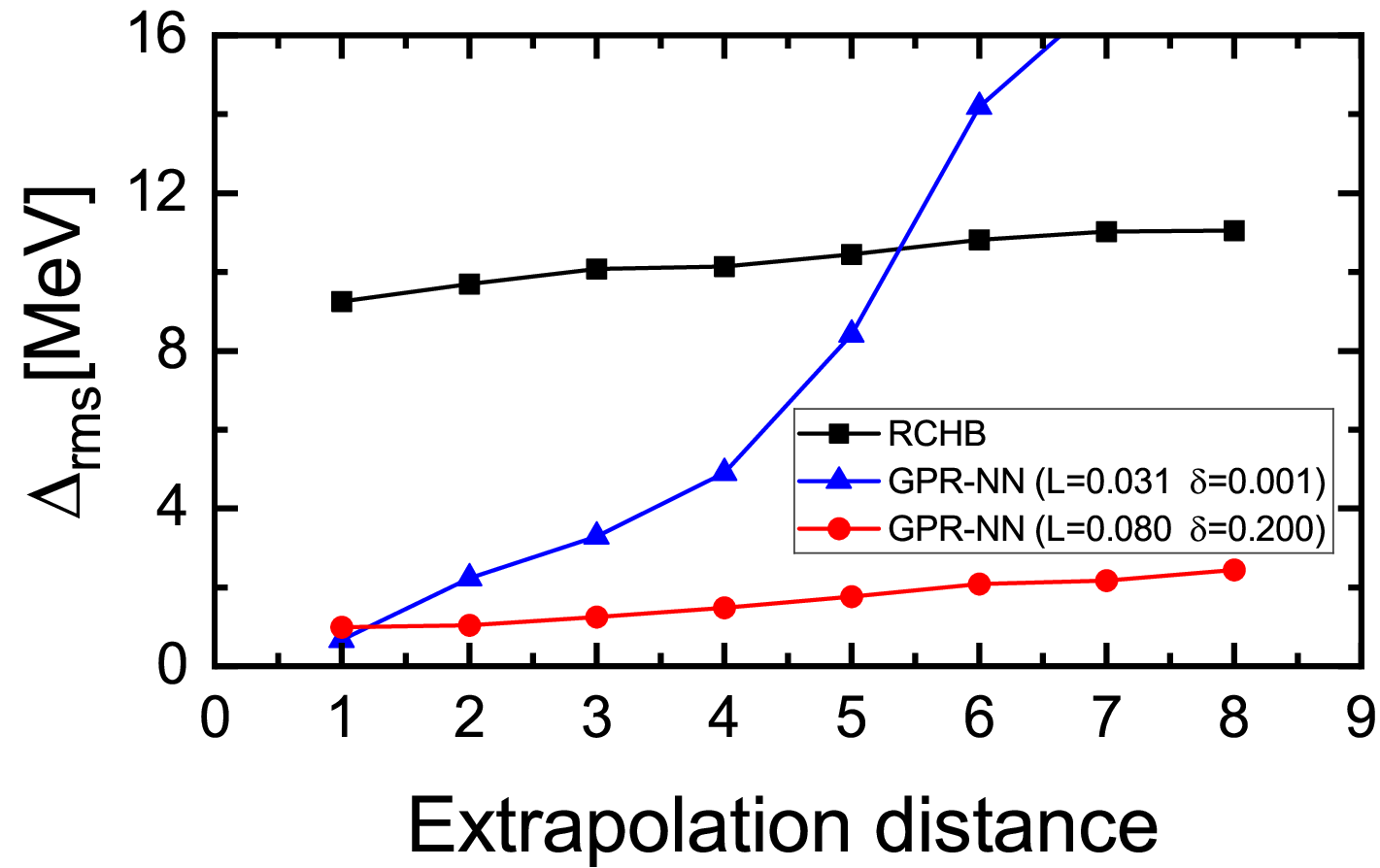}
  \caption{Comparison of the extrapolation power of the RCHB mass model, the GPR-NN approach with hyperparameters optimized in the interpolation (i.e., $L=0.031, \delta=0.001$), and the GPR-NN approach with a new set of hyperparameters (i.e., $L=0.080, \delta=0.200$) for eight test sets with different extrapolation distances (see text for details). }
  \label{fig4}
\end{figure}

In Fig.~\ref{fig4}, the rms deviations $\Delta_{\rm rms}$ of the calculated masses for the eight test sets from the RCHB mass model, the GPR-NN approach with hyperparameters optimized in the interpolation (i.e., $L=0.031, \delta=0.001$), and the GPR-NN approach with a new set of hyperparameters (i.e., $L=0.080, \delta=0.200$),
with respect to the experimental masses are shown as functions of the extrapolation distance.

One can see from Fig.~\ref{fig4} that when the set of hyperparameters ($L=0.031, \delta=0.001$) optimized in the interpolation is adopted, the GPR-NN approach does not perform well in the extrapolation.
The rms deviations $\Delta_{\rm rms}$ increase rapidly with the extrapolation distance, and they are even larger than the ones for the RCHB mass model, which means that the GPR-NN worsens the RCHB prediction instead of improving it in such extrapolation distance.
This at least indicates that the hyperparameters optimized in the interpolation through ten-fold cross-validation are not guaranteed to perform well in the extrapolation. Optimal $L$ in the training region was low due to high corrugation of the mass as a function of $Z$ and $N$, and the low $L$ naturally is detrimental to extrapolation.

However, the GPR-NN approach can perform very well in the extrapolations if one uses other sets of hyperparameters, e.g., ($L=0.080, \delta=0.200$), a new set of hyperparameters obtained by optimizing the GPR-NN predictions for these eight extrapolation test sets.
Note that the new set of hyperparameters has larger length scale of the kernel $L$, which is important for the extrapolation.
It also has larger regularization parameter $\delta$, which helps reduce overfitting.
As can be seen from Fig.~\ref{fig4}, with new hyperparameters optimized for extrapolation, the GPR-NN can significantly improve the RCHB predictions even at large extrapolation distances.
This indicates that the GPR-NN approach loses its extrapolation capability relatively gradually as the extrapolation distance increases, which is an important feature for studying $r$-process-related neutron-rich nuclei far from the experimentally known region.

\begin{figure}[!htbp]
  \centering
  \includegraphics[width=0.45\textwidth]{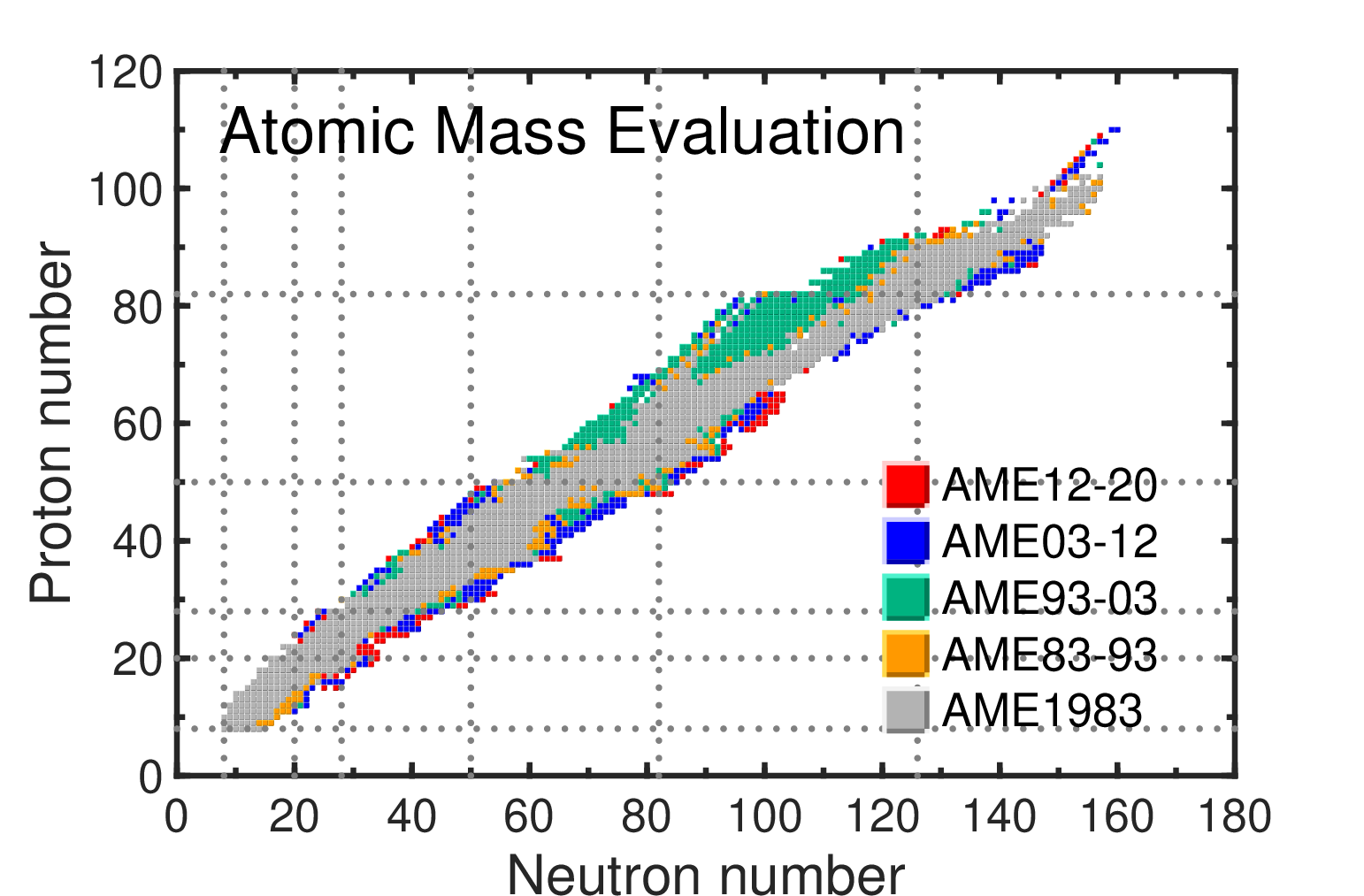}
  \caption{The nuclear landscape for nuclei with the masses experimentally measured.
  The nuclei with masses firstly compiled in different time periods for AME, including AME1983 \cite{Wapstra1985NPA}, AME83-93 \cite{Audi1993NPA}, AME93-03 \cite{Audi2003NPA}, AME03-12 \cite{Wang2012Chin.phys.C}, and AME12-20 \cite{Wang2021Chin.Phys.C}, are labeled by different colors.}
	\label{fig5}
\end{figure}

In order to study the generalization ability of the GPR-NN approach, the available 2278 data are divided into five sets according to the corresponding releasing time of the AME series~\cite{Wapstra1985NPA,Audi1993NPA,Audi2003NPA,Wang2012Chin.phys.C,Wang2021Chin.Phys.C}.
The details of the division are given in Fig.~\ref{fig5}.
The nuclei in the gray part are the overlap between the AME2020~\cite{Wang2021Chin.Phys.C} and AME1983~\cite{Wapstra1985NPA}, except for those with errors beyond 100 keV.
This gray part, labeled as AME1983 in the following, includes 1335 nuclei, and is considered as the training set.
The other four groups, AME83-93 with 177 nuclei, AME93-03 with 404 nuclei, AME03-12 with 254 nuclei, and AME12-20 with 108 nuclei, are taken as the test sets, which correspond to the new data in AME1993, AME2003, AME2012, and AME2020, respectively.

\begin{figure}[!htbp]
  \centering
  \includegraphics[width=0.45\textwidth]{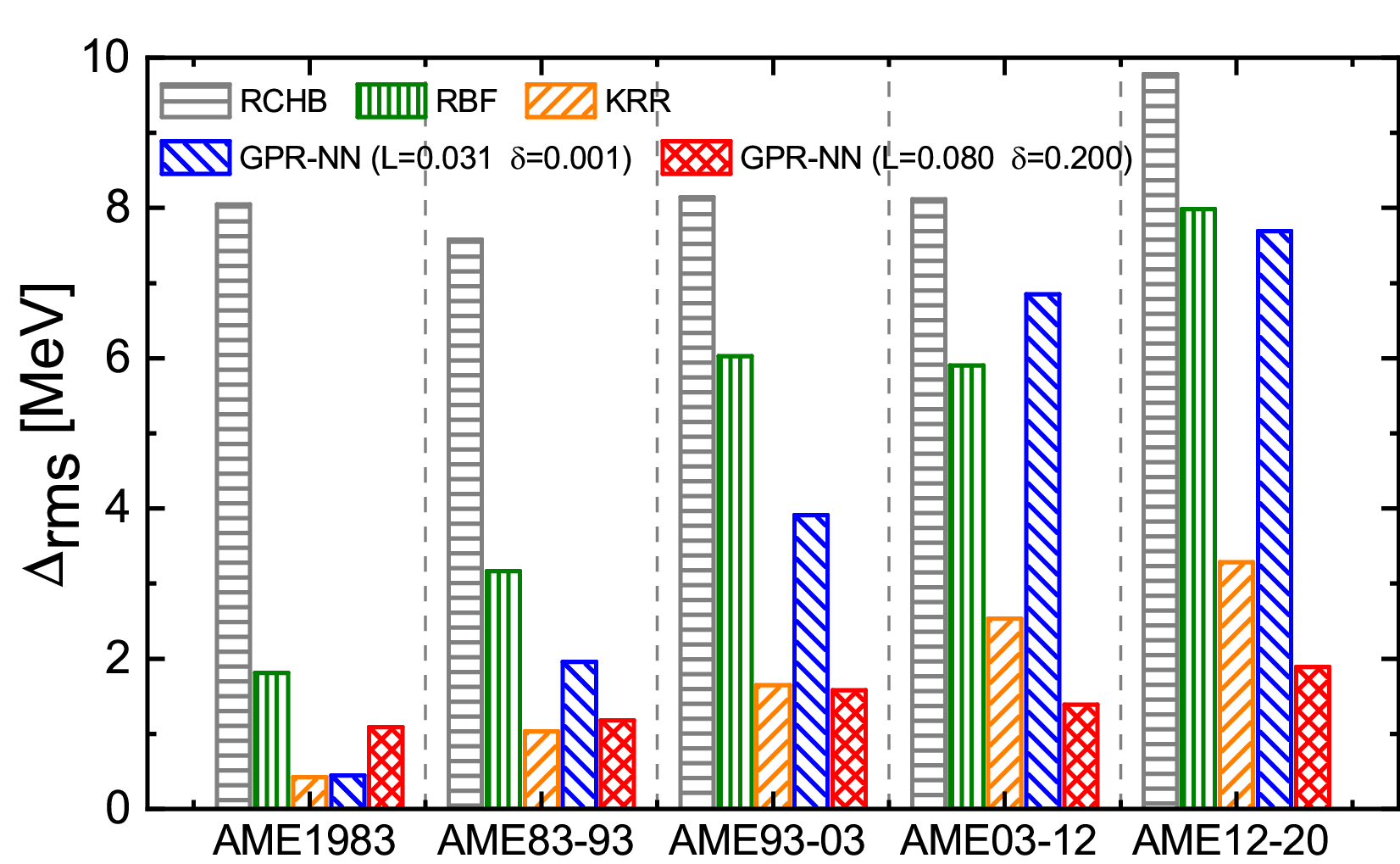}
  \caption{The rms deviations of nuclear mass $M$ of the RCHB, RBF, KRR, and GPR-NN predictions from the available data in the 5 sets, as divided in Fig.~\ref{fig5}.
  Two sets of hyperparameters, optimized for interpolation and extrapolation respectively, are adopted for the GPR-NN approach.
  }
  \label{fig6}
\end{figure}

Figure \ref{fig6} presents the $\Delta_{\rm rms}$ of nuclear mass in the RCHB, RBF, KRR, and GPR-NN predictions relative to the the five sets of available data.
Two sets of hyperparameters, optimized for interpolation and extrapolation respectively, are adopted for the GPR-NN approach.
The RBF and KRR predictions, obtained using the same manner as in Ref.~\cite{Guo2024PRC}, are provided for comparison.
In the predictions for set AME1983, the leave-one-out cross-validation is applied.
In the predictions for sets AME83-93, AME93-03, AME03-12, and AME12-20, predictions are made using the model trained on AME1983. 
One can see that the GPR-NN with hyperparameters optimized for interpolation achieves better performance than that with hyperparameters optimized for extrapolation. 
This is reasonable, as leave-one-out predictions for set AME1983 are certainly related to interpolation.
One can see that the GPR-NN with hyperparameters optimized for extrapolation always provide a much lower $\Delta_{\rm{rms}}$ in all other four sets of data.

It should be noted that the GPR-NN with hyperparameters optimized for extrapolation achieves better performance than the RBF and KRR approaches for large extrapolation sets, i.e., AME03-12 and AME12-20.
The $\Delta_{\rm{rms}}$ of the GPR-NN approach increases slowly with the extrapolation distance (from AME83-93 to AME12-20).
Significant improvements from the GPR-NN corrections can still be observed even for the AME12-20 set.
This means that the GPR-NN approach trained with the mass data released in AME1983~\cite{Wapstra1985NPA} can still help improve the predictions of masses that became available in experiments more than 30 years later. 
In other words, the GPR-NN approach trained with the mass data released in 2020 \cite{Wang2021Chin.Phys.C} would potentially help improve the mass predictions of a large number of currently experimentally unknown nuclei.

\section{Summary}

In summary, we have applied the neural network with additive Gaussian process-optimized activation functions (GPR-NN) to enhance nuclear mass predictions for the first time.
By combining the strengths of neural networks and Gaussian processes, GPR-NN effectively models nuclear mass residuals relative to the RCHB theoretical predictions using experimental data from AME2020 for 2278 nuclei with $Z\geq 8$ and $N\geq 8$.
Hyperparameters, including the number of redundant coordinates $R$, kernel length scale $L$, and regularization parameter $\delta$, are optimized through tenfold cross-validation for interpolation, yielding $R = 3$, $L = 0.031$, and $\delta = 0.001$, with an rms deviation of approximately 440 keV. 
For extrapolation, a distinct set $(L = 0.080, \delta = 0.200)$ is found optimal,  with rms deviations increasing gradually even at large distances.
Generalization tests using historical AME data sets (1983-2020) indicate that GPR-NN performs well in long-range extrapolations, maintaining significant improvements for nuclei measured decades later. 
Overall, this study validates the GPR-NN as a robust and promising tool for nuclear mass predictions. 
Its good extrapolation performance, when paired with appropriate hyperparameters, makes it valuable for improving the mass predictions for experimentally unknown, neutron-rich nuclei.
This lays the groundwork for future applications in nuclear physics and nuclear astrophysics, such as refining $r$-process nucleosynthesis models.

\begin{acknowledgments}
This work was partly supported by the National Natural Science Foundation of China under Grant No. 12405134 and the start-up grant XRC-23103 of Fuzhou University.
\end{acknowledgments}

\bibliography{paper}

\end{document}